\documentclass[sigconf]{acmart}

\def\BibTeX{{\rm B\kern-.05em{\sc i\kern-.025em b}\kern-.08emT\kern-.1667em\lower.7ex\hbox{E}\kern-.125emX}}

\copyrightyear{2019}
\acmYear{2019}
\setcopyright{acmcopyright}
\acmConference[ACSAC '19]{2019 Annual Computer Security Applications Conference}{December 9--13, 2019}{San Juan, PR, USA}
\acmBooktitle{2019 Annual Computer Security Applications Conference (ACSAC '19), December 9--13, 2019, San Juan, PR, USA}
\acmPrice{15.00}
\acmDOI{10.1145/3359789.3359841}
\acmISBN{978-1-4503-7628-0/19/12}

\iffalse
\newcommand{\luisnew}[1]{\textcolor{blue}{#1}}
\else
\newcommand{\luisnew}[1]{{#1}}
\fi

\iffalse
\newcommand{\akkaya}[1]{\authnote{KA}{#1}}
\newcommand{\leo}[1]{\authnote{Leo}{#1}}
\newcommand{\su}[1]{\authnote{su}{#1}}
\newcommand{\Luis}[1]{\authnote{Luis}{#1}}
\else
\newcommand{\akkaya}[1]{}
\newcommand{\leo}[1]{}
\newcommand{\su}[1]{}
\newcommand{\Luis}[1]{}
\fi

\usepackage{blindtext}
\usepackage{tikz}

\newcommand*\circled[1]{\tikz[baseline=(char.base)]{
            \node[shape=circle,fill,inner sep=1pt] (char) {\textcolor{white}{#1}};}}

\newcommand{\pie}[1]{%
\begin{tikzpicture}
 \draw (0,0) circle (1ex);\fill (1ex,0) arc (0:#1:1ex) -- (0,0) -- cycle;
\end{tikzpicture}%
}

\newcommand{\name}{{\textsc{\small{HDMI-Walk}}}\xspace}
\newcommand{\namehuge}{{\textsc{HDMI-Walk}}\xspace}

\newcommand{\attackone}{Attack 1 - Topology Inference Attack (Local and Remote).}
\newcommand{\attacktwo}{Attack 2 - CEC-Based Eavesdropping (Local).}
\newcommand{\attackthree}{Attack 3 - WPA/WPA2 Handshake Theft (Local).}
\newcommand{\attackfour}{Attack 4 - Targeted Device Attack (Local and Remote).}
\newcommand{\attackfive}{Attack 5 - Display Broadcast DoS (Local and Remote).}
\newcommand{\CEC}{CEC }

\begin{document}

\title[\namehuge: Attacking HDMI Distribution Networks via CEC Protocol]{\namehuge: Attacking HDMI Distribution Networks via Consumer Electronic Control Protocol}

\author{Luis Puche Rondon, Leonardo Babun, Kemal Akkaya, and A. Selcuk Uluagac}
\affiliation{%
  \institution{Cyber-Physical Systems Security Lab\\
  Department of Electrical and Computer Engineering\\
  Florida International University, Miami, Florida, USA
  }
}
\email{Email: {lpuch002, lbabu002, kakkaya, suluagac }@fiu.edu}

\keywords{HDMI Distributions, CEC, Consumer Electronics Control, DoS, Cyberattacks}

\ccsdesc[500]{Security and privacy~Denial-of-service attacks}
\ccsdesc[500]{Security and privacy~Distributed systems security}
\ccsdesc[300]{Security and privacy~Denial-of-service attacks}

\begin{abstract}
The High Definition Multimedia Interface (HDMI) is the backbone and the de-facto standard for Audio/Video interfacing between video-enabled devices. Today, almost tens of billions of HDMI devices exist in the world and are widely used to distribute A/V signals in smart homes, offices, concert halls, and sporting events making HDMI one of the most highly deployed systems in the world. An important component in HDMI is the Consumer Electronics Control (CEC) protocol, which allows for the interaction between devices within an HDMI distribution network. Nonetheless, existing network security mechanisms only protect traditional networking components, leaving \CEC outside of their scope. In this work, we identify and tap into CEC protocol vulnerabilities, using them to implement realistic proof-of-work attacks on HDMI distribution networks. We study, how current insecure CEC protocol practices and carelessly implemented HDMI distributions may grant an adversary a novel attack surface for HDMI devices otherwise thought to be unreachable through traditional means. To introduce this novel attack surface, in this paper, we present \name, which opens a realm of remote and local \CEC attacks to HDMI devices. Specifically, with \name, an attacker can perform malicious analysis of devices, eavesdropping, Denial of Service attacks, targeted device attacks, and even facilitate other well-known existing attacks through HDMI. With \name, we prove that it is feasible for an attacker to gain arbitrary control of HDMI devices. We demonstrate the implementations of both local and remote attacks with commodity HDMI devices including Smart TVs and Media Players. Our work aims to uncover vulnerabilities in a very well deployed system like HDMI distributions. The consequences of which can largely impact HDMI users as well as other systems which depend on these distributions. Finally, we discuss security mechanisms to provide impactful and comprehensive security evaluation to these real-world systems while guaranteeing deployability and providing minimal overhead, while considering the current limitations of the CEC protocol. To the best of our knowledge, this is the first work solely investigating the security of HDMI device distribution networks.
\end{abstract}

\maketitle
\section{Introduction}
\label{sec:Intro}


Audio/Video(A/V) devices have always witnessed a wide range of adoption as consumer electronics. The High Definition Multimedia Interface (HDMI) is used primarily for the distribution of A/V signals and has become the de-facto standard for this purpose \cite{hdmistandard}. For instance, in many applications such as concert halls or sporting events, large displays are chained together via HDMI to show concert images and gameplay. Indeed, as of this writing, there have been close to 10 billion HDMI devices distributed, making \luisnew{HDMI one of the most highly deployed systems worldwide\cite{HDMIBil2}}. With the requirement to merge control and communication over a single connection, the HDMI Consumer Electronics Control (CEC) protocol was specified with the release of the HDMI v1.2a \cite{HDMI}. \CEC provides control and communication between HDMI devices through HDMI cabling. This has led many vendors to implement \CEC features on their devices under different trade names, including: Anynet+ (Samsung), Aquos Link (Sharp), BRAVIA Link/Sync (Sony), CEC (Hitachi), CE-Link and Regza Link (Toshiba), SimpLink (LG), VIERA Link (Panasonic), EasyLink (Philips), Realink (Mitsubishi) \cite{tradenames}. The adoption of \CEC has become a means of control for well-known household devices (e.g., Google Chromecast, Apple TV, Sony A/V Receivers, Televisions). This rapid adoption has made \CEC into an ubiquitous protocol in many A/V installations and the adoption of \CEC enabled devices in conference rooms, homes, offices, government, and secure facilities. Given the popularity and the penetration of HDMI-based devices, their security is of utmost importance.

\begin{figure}[t]
\centering{\includegraphics[width=0.35 \textwidth]{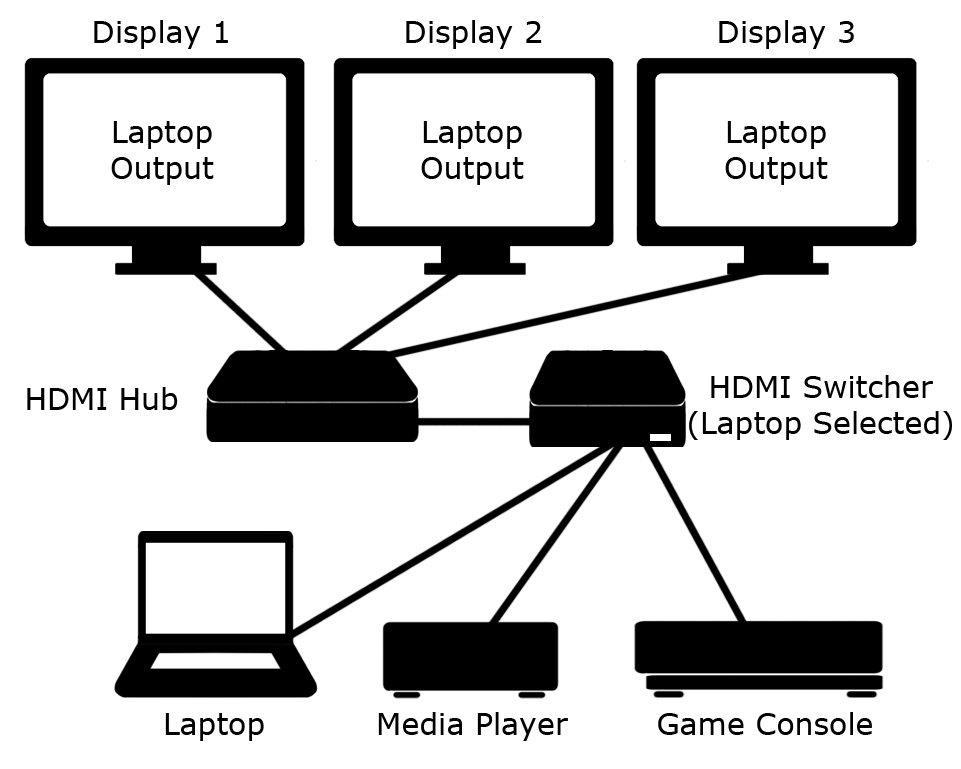}}
    \caption{Example HDMI device distribution network including three displays sharing the same source image (Laptop). Usually, in bars and conference rooms, displays are chained via HDMI cables.}
\label{hdmidistribution}
\vspace{-0.3in}
\end{figure}

Nonetheless, \CEC is outside the reach of the traditional networking mechanisms, and most importantly, current security mechanisms provide no protection to CEC-based threats. This creates a widely-available, unprotected, and unexplored threat vector in locations (e.g., homes, government, offices, etc.) without mainstream user awareness. Unprotected HDMI networks give malicious entities an attractive medium of attack from which they can remain undetected. \CEC allows them to perform activities over an HDMI device distribution network such as information gathering, device control, and attack facilitation. In effect, an attacker can retrieve and alter the power state of all HDMI-Capable devices without physical or traditional network access. While there has been abundant research on the security of traditional networks, this protocol has remained an under-researched communication component in the realm of cybersecurity. 

\textit{Contributions:} In this work, we introduce \name as a novel attack vector to deployed HDMI distribution networks. We present five types of both local and remote attacks involving HDMI devices. We evaluate how \CEC can be exploited to an attacker's advantage and an adversary can implement these attacks. We demonstrate the threat of \name with a specific testbed of commonly used HDMI equipment (e.g., Google Chromecast and Sharp Smart TV) for the evaluation of \name attacks. Specifically, as proof-of-concepts, we performed five novel \name attacks involving: (1)  malicious topology inference; (2) Denial-of-Service attacks; (3) audio eavesdropping with sensitive data transfer over HDMI; (4) targeted device attacks to disrupt services through HDMI; and (5) finally the facilitation of existing wireless-based attacks with a CEC-enabled HDMI device. We also evaluated the implications of these attacks as part of this work. With the execution of these novel attacks, we prove that the arbitrary control of \CEC devices is feasible for an attacker. 

This work aims to uncover vulnerabilities in HDMI distributions, a very well-deployed, ubiquitous system. Consequences of these vulnerabilities provide impactful security issues to HDMI users and systems dependent on HDMI. Ultimately, opening discussion on defense mechanisms and impactful security practices specific to CEC while considering CEC protocol limitations. To the best of our knowledge, this is the first work that solely investigates the security of \name-styled attacks through HDMI distribution networks. 

\textit{Organization:} The rest of the paper is organized as follows. Section II presents some background information on HDMI device distributions and the \CEC protocol. Section III covers the related work on \CEC security. Section IV presents the assumptions, definitions and HDMI threat model in our paper. In Section V, we cover the architecture for the novel \name attacks. In Section VI, we describe our testbed, software, modules, attack implementations and evaluate the findings and implications of \name attacks. In Section VII, we discuss possible mitigation strategies and defense mechanisms. Finally, we conclude this paper in Section VIII.

\section{Background} \label{sec:backgroud}

In this section, we present some necessary concepts about the Consumer Electronics Control (CEC) protocol and distributed HDMI-based device setups.

\subsection{The High Definition Multimedia Interface (HDMI)}

The High Definition Multimedia Interface or HDMI, was developed with the purpose of digital Audio/Video transfer with seamless integration of communication features through the same connection \cite{HDMISpec}. Through the 19-pin connector, HDMI transfers Audio, Video, Network, and CEC communication signals \cite{HDMIPinout}. With almost ten billion HDMI-capable devices distributed around the globe, HDMI has found a place in countless residences, offices and secure facilities and has become one of the most widely-deployed protocols worldwide \cite{HDMIBil2}. In its current form, HDMI is primarily used in high-bandwidth, video distribution applications between vastly different types of devices (e.g., Televisions, Bluerays, Media Centers, etc.).

\begin{table}[t]
    \centering
    \caption{\footnotesize{\CEC Logical Address Assignment. \cite{HDMIBUS}}}
    \begin{tabular}{|c|c|}
    \hline
       Logical Address & Device Type\\ 
       \hline
       0 & Television\\
       1,2 & Recording Devices\\
       3,6,7,A & Tuners\\
       4,8,9,B & Playback Devices\\
       C,D & Reserved\\
       E & Free Use \\
       F & Broadcast\\
    \hline
    \end{tabular}
    \label{idassign}
    \vspace{-0.2in}
\end{table}

\subsection{HDMI Distribution Networks}
HDMI deployments are not limited to one-to-one connections. Similar to Ethernet networks, there are many devices which control the HDMI signal flow and distribute signal in a controlled and organized manner. For instance, in Figure \ref{hdmidistribution} the user maintains the same visual image over three displays, and switches between three source devices. This figure also shows the laptop selected as the active source over multiple displays. Depending on the device setup, there is a distribution of \CEC through the same connection. We note the following components in an HDMI distribution and will refer to them during our work.

\noindent\textit{Displays:} Any device with a primary purpose of being an end-display such as a television or a projector.
   
\noindent\textit{Hubs/Splitters:} Any device which primarily allows multiple video signals to be split to various displays from a single video input without switching.
   
\noindent\textit{Switches:} Any device with a primary purpose of allowing various source device inputs to one or more display device outputs. They also perform switching between these sources to a different output(s).

\noindent\textit{Source Devices:} Any device which is primarily an HDMI output-only devices such as a Chromecast or a laptop.
   
\vspace{-0.00in}
\subsection{The Consumer Electronics Control (CEC) Protocol}

\CEC was developed in to enable interoperability between HDMI devices, with full specification in 2005 \cite{HDMI}. \CEC signals are carried through Pin 13 as part of the HDMI interface \cite{HDMIPinout}. The communications in \CEC are divided into 10-bit blocks that include a header, opcode, and data blocks. The flow of information is dictated by the header, the first eight bits note the source and destination. Message Destination may refer to a specific device by logical address or broadcast as noted in Table \ref{idassign}. This broadcast functionality is especially exploited by \name attacks.

\begin{figure}[t]
\centering{\includegraphics[width=0.43 \textwidth]{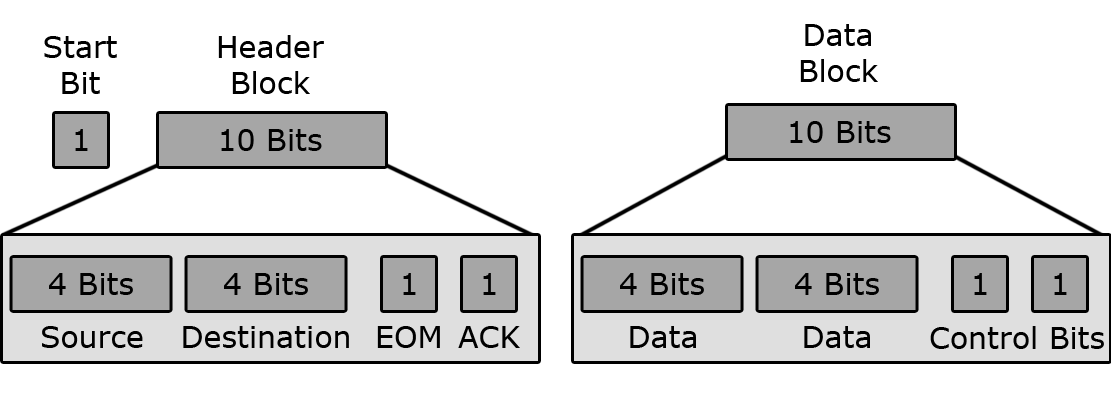}}
    \caption{\footnotesize{The \CEC stack and structure as used in HDMI}}
\label{cecdata}
\vspace{-0.1in}
\end{figure}

Figure \ref{cecdata} shows how the \CEC header allows for 16 unique IDs (4 bits). IDs 0-E specify device addresses while the last logical address (F) is reserved for broadcast 
within the HDMI distribution. This logical address assignment usually follows certain device-type guidelines. For example, displays are usually assigned to the logical address (0) and additional displays self assign to ``free use'' (E) as shown in Table \ref{idassign}.

\section{Related Work} \label{sec:relatedwork}
\vspace{-0.05in}

There has been some research in compromising A/V devices through a variety of attacks. Work from Zhang et al., presented a security overview on connected devices and noted common vulnerabilities such as weak authentication, over-privilege and implementation flaws in connected devices \cite{ZhangIoT}.  Within the scope of Smart TVs, Oren \& Keromytis describe a method of compromising connected Smart TVs through Hybrid Broadcast-Broadband Television (HbbTV) and web-based code injections \cite{OrenKeromytis}. Related work from Niemietz et al., on Smart TVs explores the attacks on Smart TVs through app-based approach \cite{NiemSmartTV}. This work centers on TV embedded applications, and the security flaws which may come from vendor-specific apps. 
On the other hand, research related to HDMI systems and their security issues is relatively an uninvestigated avenue or not systematically investigated yet by the research community.  The most relevant work in HDMI systems is a 2012 work published by NCC Group, which focused on vulnerabilities with fuzzing~\cite{nccgroup}. Similarly, Smith presented \CEC as an avenue of attacks through fuzzing \cite{smithcec}. And, further work presented CECSTeR as a fuzzing tool \cite{davisblackhat}.

\textit{Our work differs from these works as follows:
We introduce a novel attack method called \name to HDMI devices.  Our scope is entirely through \CEC as the main vector of attack and does not rely on any custom applications, software vulnerabilities, fuzzing, buffer-overflows,  vendor-specific attacks, or traditional network connectivity. We focus on the exploitation of the \CEC protocol in both local and remote attacks. We demonstrate proof-of-concept implementations of five different types of attacks; specifically, (1) malicious device Scanning, (2) eavesdropping, (3) facilitation of attacks(e.g,. WPA Handshake theft), (4) information theft, and (5) denial of service through HDMI.}

\begin{figure*}[!th]
\centering{\includegraphics[width=0.7\textwidth]{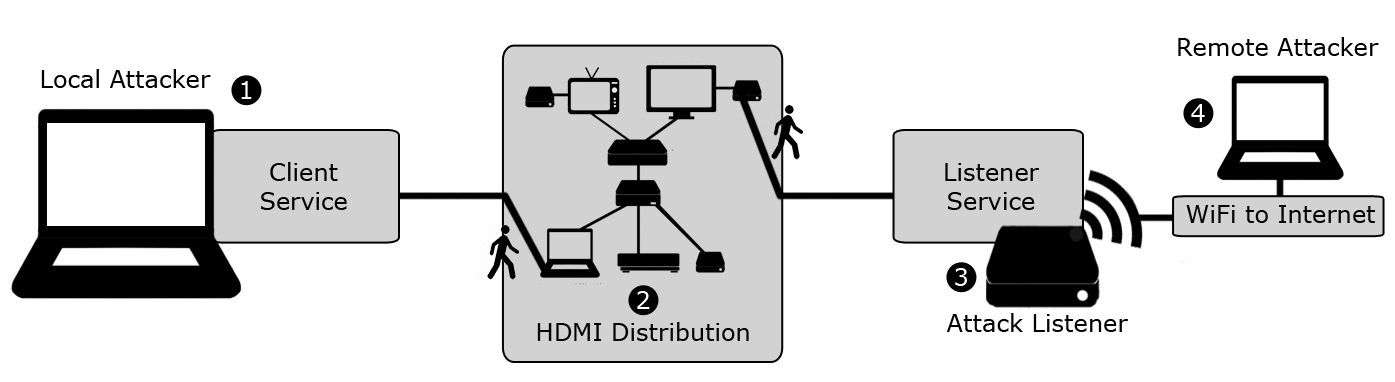}}
    \caption{General architecture for \name-based attacks.}
\label{fig:architecture}
\vspace{-0.15in}
\end{figure*}

\section{Problem, Assumptions, and Threat Model}

In this section, we present the assumptions, definitions, and the threat model for \name-based attacks.

\vspace{-0.1in}
\subsection{Problem Scope}
This work denotes an HDMI distribution network within a conference room which may be used for confidential presentations. The topology of this distribution network includes common HDMI distribution equipment such as switches, hubs as well as HDMI devices such as displays and sources. The attacker is an invited guest presenter Mallory, who has a small amount of time to prepare in the conference room without any supervision. Mallory either compromises an existing HDMI device through malware, or hides a malicious HDMI-capable device within the distribution (e.g., connected behind a television). Mallory connects her own laptop to auxiliary ports on the podium prior and during the presentation and perpetrates the \name attacks. After presenting, Mallory leaves. Sometime after her departure, further security policies are enacted and unsupervised access to the conference room is disallowed to visitors. Mallory's only avenue of attack is to access her hidden device indirectly, locally or remotely.

\textit{Compromising Devices:} We note that Mallory may compromise an HDMI distribution without direct access to the HDMI network. Malware (e.g., firmware, app-based) could compromise an existing device to Mallory's benefit, acting as a link between the distribution and their machine. For instance, privileged malware applications in an Android-based A/V device could make use of the \texttt{HdmiControlManager} functionality which is available to transmit and receive arbitrary messages \cite{AndroidCEC}.  

\textit{Possible Payloads:} CEC attacks can provide access to devices which may have been believed secure or isolated in a conference room. Conference rooms may serve varying purposes from unrestricted to confidential usage. When a space is in unrestricted usage, an attacker may disrupt operation, damage equipment and prevent normal usage of a conference room through CEC attacks. If the space is used for confidential purposes, an attacker may gather data about a system, gather restricted information within a conference room, or simply facilitate more complex attacks (e.g., wireless handshake theft, eavesdropping). In both of these cases, an attacker may avoid traditional means of detection through the use of CEC.

\textit{Attack Mode 1 (Local Communication): } Mallory only has local access when connecting directly to the HDMI distribution network as a presenter. This case is independent of any form of network access, it relies on Mallory's ability to connect to the auxiliary connection on the conference room podium. Local communication from her laptop through the HDMI distribution with \name and to the hidden or compromised device.

\textit{Attack Mode 2 (Remote Communication): } In this case, Mallory has found an open guest network connection during her first visit or later gained unauthorized internet access. This allows Mallory to enable remote access to her hidden device. Furthermore, this allows Mallory to perform specific attacks.

\vspace{-0.10in}
\subsection{Definitions}

In this sub-section we denote definitions for concepts used in \name. 

\noindent\textit{Definition 1 - Isolated Device: } An HDMI device which has no network connectivity to traditional IP networks in any manner. 

\noindent\textit{Definition 2 - Limited Access User: } A limited-access user is primarily described as a user with temporary physical access to a location and limited IP network connectivity. This user can be a temporary visitor such as a presenter. 

\noindent\textit{Definition 3 - Attacker (Temporary Visitors): } An attacker is any limited-access user which attempts malicious access to unauthorized resources. The attacker's motivations are to disrupt, gather information, gain unauthorized access, learn user behavior, and perpetrate the attacks listed in the threat model below. In our case, the attacker may be a temporary visitor with limited access to the facilities (e.g., a presenter, Mallory). 

\vspace{-0.1in}

\subsection{Assumptions}

To perform the \name attacks, we have the following assumptions.

\noindent\textit{CEC Propagation:} This work assumes full \CEC protocol propagation over the distribution of HDMI devices. Some devices tested had no function to disable CEC propagation, even if CEC control was disabled. In testing performed on devices with multiple HDMI ports, we found 80\% of devices provided some form of propagation.     

\noindent\textit{CEC Control:} We assume CEC control is active on connected devices in the distribution. This is a realistic scenario, as we found that in all CEC-capable devices tested, CEC functionality was enabled by default. We also observed that many devices revert to default settings after a firmware update.
   
\noindent\textit{Access to HDMI Components:} We also assume that Mallory has access to some HDMI components (or endpoints) in the distribution. This is a realistic assumption as A/V components are often not as secure as networking components. Display inputs and outputs are often visible and available to presenters. Presenters are often given enough time to prepare and free access to A/V equipment in a conference room without supervision or suspicion is expected. \luisnew{In some cases, we have found displays (often used for information purposes) outside conference rooms which could act as another connection point to an HDMI distribution inside a conference room.}

\vspace{-0.1in}

\subsection{Threat Model}

\name assumes the following five threats as part of the threat model. 

\textit{Threat 1: Malicious \CEC Scanning:} This threat considers the malicious use of \textit{scanning} features through \CEC and exposed HDMI ports to gather information about the connected devices. For instance, Mallory can create a topology of available HDMI devices to control and use this information to perform further attacks.

\textit{Threat 2: Eavesdropping:} In this threat, Mallory is not present but actively eavesdrops on \CEC communication through an implanted device. 

\textit{Threat 3: Facilitation of attacks:} This threat eliminates time and physical access limitations in wired and wireless attacks. \name facilitates many of these attacks so that they become more viable or more difficult to detect. For example, Mallory installs a device to passively capture WPA handshakes, avoid detection, and control through \CEC remotely.

\textit{Threat 4: Information Theft:} This threat considers information theft as a form of data transfer which Mallory may find valuable. For example, information about available HDMI devices or wireless handshake capture which would enable future attacks.

\textit{Threat 5: Denial of Service:} This threat considers Denial-of-Service attacks where Mallory disrupts the availability of a system through an HDMI connection. These attacks may be targeted to a specific device or broadcast to multiple devices. For example, Mallory prevents the use of a television through the repeated broadcast of HDMI control commands.

Note that this work does not consider attacks which focus entirely on IP networks; data injection attacks through \CEC such as buffer overflows over \CEC or setting manipulation attacks. Similarly, other protocols such as USB or Bluetooth are entirely outside the scope of this paper.

\vspace{-0.1in}
\section{\textsc{HDMI-Walk}}\label{sec:architecture}
In this section, we present the details of the \name based attacks.  Figure \ref{fig:architecture} depicts the general architecture of \name which comes with four main components: \textit{local attacker, HDMI Distribution, attack listener, and remote attacker.} 

The first component of HDMI-Walk is the Local Attacker which runs the Client Service in their local machine. This local hardware is temporarily connected to the HDMI distribution. The client service contains any required modules for communication to the listener and facilitates the attacks through \name (\circled{1}). The second part is the HDMI Distribution, which is the core of our attacks and allows for end-to-end communication between devices through HDMI as a medium. The user may scan the distribution for addressed CEC devices, as well as communicate bidirectionally with other devices (\circled{2}). The third part of the architecture involves the Attack Listener. The attack listener is the physical attacker device and hosts the Listener Service. The listener service includes all the required modules for \name communication and listener-run attacks. This service also includes a remote access module to enable communication to the remote client if a connection is available (\circled{3}). Finally, we have the Remote Attacker, which communicates directly through a remote connection to the attack listener if remote access is possible (\circled{4}). 

\noindent \textbf{Local Attacker:} A local attacker establishes communication with the listener device through \CEC and the HDMI Distribution. 
The local attacker places their client device in an exposed HDMI port such as an auxiliary connection in a presentation room or a side input of a television. In our case, the client device can be a laptop with a \CEC capable adapter. The client's main purpose is to establish communication with the listener and serves as the main interface for an attacker to issue commands and receive data from the listener device.  The local client communicates to the listener through \name derived control of the distribution. Additionally, the client device hosts the client service. This service contains all necessary software modules for specific actions within the scope of \CEC such as the ability for file transfer, arbitrary \CEC communication, and \CEC scanning. 

\noindent \textbf{HDMI Distribution:} This allows for the core concepts of this work is the nature of the CEC protocol which allows propagation and control. These are not inherently equal or mutually exclusive; for instance, a device may be able to both control and propagate \CEC commands through auxiliary HDMI ports. In contrast, a different device, such as an HDMI hub, may allow propagation but offer no \CEC based control. The inherent design of \CEC allows for any device to transmit and request information to and from any other device within the same distribution. During our evaluations, we found that \CEC commands propagate from device to device, passing through different 'hops' in a similar fashion to a bus network while allowing individual devices to further propagate communication to their own branched connections. This is a requirement in 'scanning' behavior, which allows for any device to query others by logical address for a name, type, language, OSD string, vendor, power status, CEC version, and source status. With this, the querying device is able to build a map of available \CEC devices within the distribution. Since the headers signify a broadcast or a message to a specific device by logical address, this becomes useful for targeting specific devices or broadcasting to all devices.

\noindent \textbf{Attacker Listener:} The listener device awaits client commands. Ideally, the listener is hidden by the attacker in a location such as behind a television, in an equipment cabinet or anywhere where there is a connection to the HDMI distribution. The listener may establish communication with CEC-enabled isolated devices (see Section IV) through \name. In the attack model, once the listener receives expected commands from the attacker client (local or remote), it will enact actions in the HDMI distribution on behalf of the attacker. In our proposed \name, the attack listener performs the core actions for our attacks and runs all the separate modules required for each attack. Additionally, the listener hosts all the software modules required by the attacker for \CEC communication, CEC file transfer, CEC scanning, microphone access, wireless access, and remote access.

\noindent \textbf{Remote Attacker:} The remote attacker maintains a remote web interface to a listener device. Commands and messages are relayed bidirectionally from the listener and the remote attacker. In contrast to the local attacker, the remote attacker operates in a remote web server, has no direct \CEC connectivity and only hosts remote communication modules. The remote attacker's server is polled via an Internet connection by the listener for new commands. This allows the attacker to perform remote execution of \CEC actions using the listener. These actions may involve \CEC information gathering,  targeted attacks, DoS, or any attack module within the listener device.

\section{Evaluation and Realization of HDMI-Walk Attacks}\label{sec:attacks}
In this section, we describe and evaluate the \name attacks in detail. The purpose of the attacker is to leverage the \name capabilities to discover, manipulate, control, and cause undesired operation to devices within an HDMI distribution. The adversary also aims to use \CEC as the primary medium for their attacks. This is achieved via a connection to the listener through a remote client or through a local HDMI connection. As explained in earlier in Sections IV and V, in all of the attacks, the attacker plants the listener device somewhere within the \CEC distribution (e.g., behind a television). 

\begin{table}[t]
\scriptsize
    \centering
    \caption{Hardware and software usage.}
    \begin{tabular}{|c|c|}
    \hline
       Hardware & Software\\ 
       \hline
       Sharp Smart TV. & Pulse Eight LibCEC 4.0.2 \\
       Samsung UN26EH4000F & Python 3.6.1\\
       Monoprice Blackbird 3x1 HDMI Switch & Aircrack-ng 1.2-rc4\\
       Wyrestorm - 1x4 HDMI 1.3b Splitter & Eclipse IDE\\
       Chromecast NC2-6A5 & PyAudio v0.2.11\\
       Sony STR-ZA2100ES & Jersey JAX-RS\\
       Raspberry 3 Model B x2 & Raspbian Version 9 \\
       TP-Link TL-WN722N V1 Adapter & Swagger.io\\
       Motorola G5 Plus Phone & Java 1.8\\
       TP-Link TL-WR841N Router & AWS Elastic Beanstalk \\
    \hline
    \end{tabular}
    \vspace{-0.2in}
    \label{testbedtbl}
\end{table}

\subsection{The implementation of \name}
\begin{figure}[t]
\centering{\includegraphics[width=0.38 \textwidth]{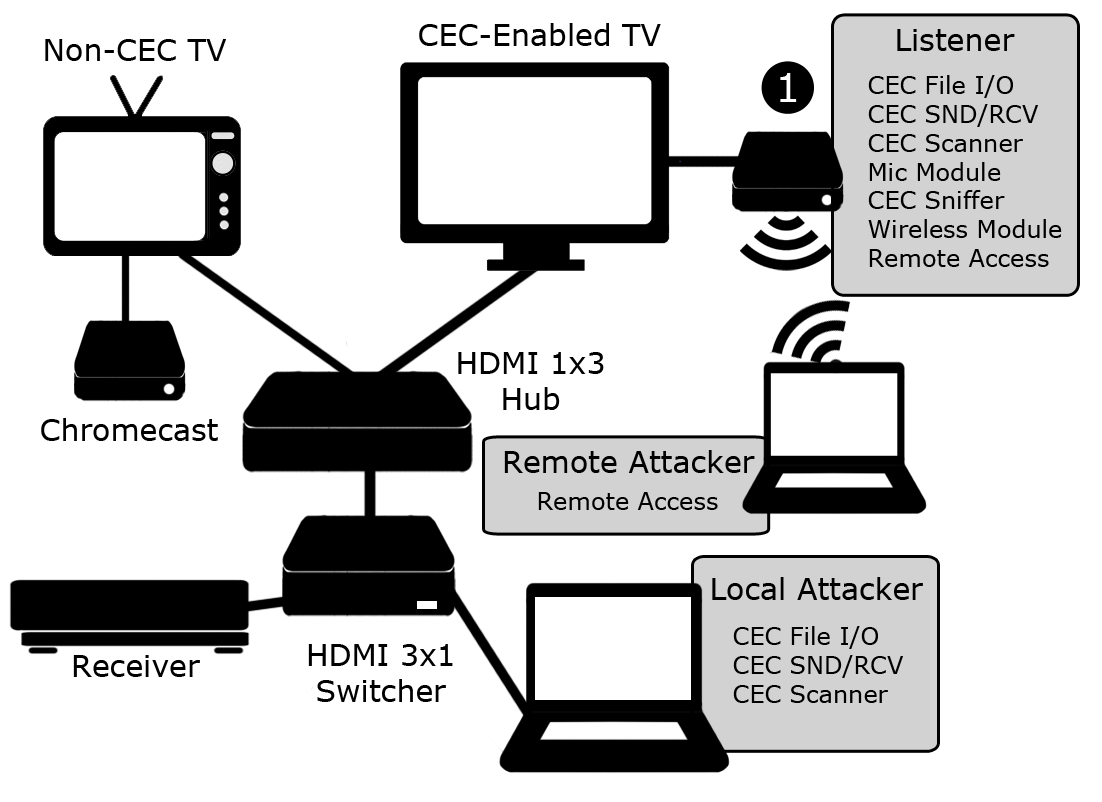}}
    \caption{\name testbed implemented with various commodity HDMI devices.}
\label{testbed}
\vspace{-0.05in}
\end{figure}

In order to ensure the attacks are implemented in a realistic HDMI environment, we created a \CEC capable testbed with standard and widely available commodity HDMI devices presented in Table \ref{testbedtbl}. Here we included two displays, an HDMI switcher, an HDMI hub, a source and the attacker devices as depicted in Figure \ref{testbed}. We utilized LibCEC, an open-source \CEC implementation \cite{libcec}. This library provides Python modules which we used to create both the client and the listener services. Due to readily available \CEC support in Raspberry Pi v3 devices, we used two Pis, one as the listener and one as the local client to perform the attacks and evaluations. To test WiFi (handshake) and remote attacks, we created a network with SSID \texttt{Portabox}. Note that even with a non-CEC-addressed TV, the TV simply propagates any \CEC commands through additional ports.

\vspace{-0.05in}
\subsection{Software Modules}

\subsubsection{CEC File I/O} This module facilitates sensitive data transfer within the \CEC protocol. We leveraged \name for data transfer between client and listener devices within the \CEC distribution. This module can be subdivided into three sections: serialization, transmission, and deserialization. We break down a file transfer from the listener (sender) to the attacker (receiver) below:
 \begin{itemize}
     \item \textit{Serialization:} LibCEC allows the transfer of \CEC packets through the distribution. The attacker device (hosting client service) first begins the file request with the ``aa:aa:aa:aa'' packet. The data is then imported into the running service and converted into hexadecimal values. This serialized file is stored locally within the buffer of the current sender (i.e., the attack listener).
     \item \textit{Transmission:}  The buffer is segmented into hexadecimal strings of length 28 in preparation for the file transfer by the sender. Each segment of the buffer is sent with the data header ``xx:00'' over \CEC to the receiver. Finally, once all segments are exhausted, the transmission ends with ``ee:ee:ee:ee''. Any packet received without these headers are dismissed by the receiving device.
     \item \textit{Deserialization:} packets are received in order by the receiver (i.e., the attacker laptop), cropped and then stored locally in a clean data buffer. With transmission finished, the client now deserializes the stored buffer into the original file.
 \end{itemize}   
\vspace{-0.08in}
\subsubsection{CEC SND/RCV} This module sends and receives custom \CEC messages through the alteration of the header (destination) and data blocks. These may be used to activate listener conditions, attacks, or request a file transfer from the listener. We achieve sending and receiving of custom \CEC messages through the use of the libCEC Python module. This library provides the communication method which allows the creation and transmission of \CEC commands within specifications. This function is used as part of File I/O transfer or to transmit specific commands to a device over the HDMI distribution.
\vspace{-0.08in}
\subsubsection{CEC Scanner} This module scans a distribution to identify \CEC devices. The CEC scanner implements the standard LibCEC \texttt{scan} command which queries all possible devices within a distribution and records all valid responses. \name captures the available devices and provides the attacker information on each device and the logical address of their listener. 
\vspace{-0.08in}
\subsubsection{Microphone Module} Used to record and store anything captured by the embedded microphone on the listener device for the purpose of audio eavesdropping. Microphone access and recording were achieved through the use of the PyAudio library. PyAudio allows local storage of audio data within a Python operation at pre-determined length and bandwidths. We created this module to activate the microphone in the listener device.
\vspace{-0.08in}
\subsubsection{CEC Sniffer} \CEC sniffer allows the listener to passively monitor all the commands and data passing through the \CEC distribution. Targeted attacks may use this feature to trigger commands upon the action of a device. This is implemented through the command callback in the LibCEC library which allowed us to handle any command received through the bus. We analyze every packet for specific calls during the attack phases. With this form of detection, attacks may target specific devices based on their power state change.
\vspace{-0.08in}
\subsubsection{Wireless Module} The wireless module comes in two parts. The first part provides standard wireless access or the capability to connect to an Internet-enabled network for remote support to the attacker. The second part implements Aircrack-ng to allow for sniffing, capture, and final cleaning of WPA/WPA2 handshakes for further cracking. We use a \texttt{monitor mode} capable adapter with this module (TP-Link) and Python calls to automate the process in the target WiFi network.

\begin{table}[t]
\scriptsize
    \centering
    \caption{Module utilization per attack} 
    Legend : \pie{180} = Local Attack, \pie{-180} = Remote Attack, \pie{0} = Neither, \pie{360} = Both \\
    \begin{tabular}{|c|c|c|c|c|c|}
    \hline 
     & Topol. & CEC & Handshk. & Targeted & Broadcast \\
     & Infer. & Eaves. & Theft & Attack & DoS\\
    \hline
    File I/O & \pie{-180} & \pie{-180} & \pie{-180} & \pie{0} & \pie{0}\\
    SND/RCV & \pie{360} & \pie{-180} & \pie{-180} & \pie{360} & \pie{360} \\
    Scanner & \pie{360} & \pie{-180} & \pie{-180} & \pie{360} & \pie{-180}\\
    Mic Mdl. & \pie{0} & \pie{-180} & \pie{0} & \pie{0} & \pie{0}\\
    CEC Sniffer & \pie{0} & \pie{0} & \pie{0} & \pie{360} & \pie{0}\\
    Wireless Mdl. & \pie{0} & \pie{0} & \pie{-180} & \pie{0} & \pie{0}\\
    Remote Mdl. & \pie{180} & \pie{0} & \pie{0} & \pie{180} & \pie{180}\\
    \hline
    \end{tabular}
    \vspace{-0.24in}
    \label{moduleuse}
\end{table}

\vspace{-0.08in}

\subsubsection{Remote Access Module} The remote access module is utilized to allow for remote requests to the listener device through a valid Internet connection. It is divided into two parts: Server Component and the Listener Web Component.

    \textit{Server Component:} We hosted a RESTful API running Swagger GUI as the remote client and server component within AWS' Elastic Beanstalk service. We create two string caches reachable with the paths \texttt{/cec/listener} and \texttt{/cec/webclient} each with GET and POST methods. The attacker accesses this server component and submits their commands through \texttt{POST: /cec/listener} with a JSON object containing the desired command to execute remotely. This server can be freely accessed through \cite{myapi}
    
    \textit{Listener Web Component: } We implemented the web component using Python threading and polling requests to our server. The listener polls \texttt{GET: /cec/listener} every two seconds for new commands submitted for remote execution. This listener component posts to \texttt{POST: /cec/webclient} for later retrieval by the attacker.

\subsection{Attacks}
In this sub-section, we realize \name attacks and discuss its implications. We also present individual uses of every module aforementioned for \name attacks in Table \ref{moduleuse}. 

\textbf{\attackone} This attack is a demonstration of Threat 1 (Malicious CEC Scanning) possible through \CEC in online and offline scenarios. We use the HDMI-Walk architecture to move through the distribution and gather information about every device available with malicious intent. This attack can be executed through the local or remote client.

\textit{Step 1 - Activation:} Upon initial placement within the HDMI distribution, the listener automatically connects and begins the information gathering process with remote and local execution of \name scans.

\textit{Step 2 - Information Gathering:} The listener begins to perform a ``walk'' over all of the devices using the CEC scanner module. This easily yields information about HDMI device type, device, logical address, physical address, active source, vendor, \CEC version, device name, and power status from available devices in the distribution. Once this has been processed, the listener stores the data locally.

\textit{Step 3 - Leakage:} For a local client, the data is ready to be retrieved through the File I/O module upon local client request. For the remote client, the listener performs a call to \texttt{POST: /cec/webclient} with all the captured information. The data is submitted to the remote server in the form of a JSON object to be retrieved by a remote attacker.

\textit{Evaluation:} With this attack, we used the scanning functionality to ``walk'' and gather more information on the controllable devices available. The attack was entirely successful and allowed us to learn information both locally and remotely about each accessible device. As seen in Table \ref{tbl:scanresults}, we gather information such as the device logical/physical address, active source state, Vendor name, CEC Version, OSD Name, and power status. With this information, an attacker may as well infer usage from the power state of the equipment. For example, an attacker may be able to infer that a room is in use when the power state of the displays is \texttt{on} or perform more vendor and device-specific attacks with more research on specific devices.

\begin{table}[t]
    \centering
    \caption{Attack 1--Information gathered through HDMI-Walk.}
    \scalebox{0.8}{
    \begin{tabular}{|c|c|c|c|c|c|}
    \hline
       Info & Addr 00 & Addr 01 & Addr 02 & Addr 04 & Addr 05 \\ 
       \hline
       P. Addr & 0.0.0.0 & f.f.f.f & 4.0.0.0 & 3.0.0.0 & 1.0.0.0 \\
       Active & No & Yes & No & No & No\\
       Vendor & Unk & Unk & Pulse-Eight & Google & Sony \\
       OSD Str & TV & RPI & CECTestr & Chromecast & STR-ZA2100 \\
       CEC Ver & 1.4 & 1.3a & 1.4 & 1.4 & 1.4\\
       Pow Status & ON & ON & ON & ON & Standby \\
       Language & Eng. & Eng. & Eng. & Unk & Unk\\
    \hline
    \end{tabular}}
    \label{tbl:scanresults}
    \vspace{-0.18in}
\end{table}

\textbf{\attacktwo} We perform this attack to demonstrate Threat 2 (Eavesdropping) and Threat 4 (Information Theft). In this local attack, an attacker has access only to the HDMI port for communication with the listener device. The attacker walks the HDMI distribution and forwards messages to the listener to activate and record audio using the Microphone Access Module. This audio data is stored locally in the listener device. The audio data is then transferred to the client at a later date through the use of the File I/O module.

\textit{Step 1 - Listener:} The attacker first places a listener device in the \CEC distribution as noted by the architecture. The listener device awaits attacker commands from another location in the HDMI distribution.

\textit{Step 2 - Listener Activation:} The attacker sends the request to performs an HDMI-walk to scan the devices and identifies the listener device in the \CEC distribution. We note the logical address of the listener device and activate the Microphone module with ``bb:bb:bb:bb'' command received by the listener. The listener device records audio and stores the data locally.

\textit{Step 3 - Client Request:} The client requests a file transfer using the File I/O module and the command ``aa:aa:aa:aa'' to the listener. The listener receives this command via the \CEC distribution and serializes the stored audio data as the client awaits the data transfer. Once the audio file is serialized the File I/O module transmission begins. 

\textit{Step 4 - Client processing:} The audio data is transmitted from the listener device to the client service through the File I/O module. Once this is finished the client saves the audio file locally, making it available to the attacker.

\begin{figure}[t]
\centering{\includegraphics[width=0.4 \textwidth]{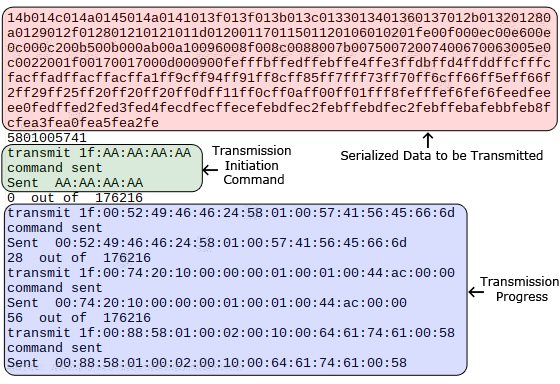}}
    \caption{Attack 2--File I/O Module transfer of audio data.}
\label{filetransfer}
\vspace{-0.2in}
\end{figure}

\textit{Evaluation: } For this attack, we had success at every stage of the attack. Tests performed in different locations of the HDMI distribution proved successful. Script activation began and a recording was saved locally. The listener device successfully received the activation command from the client and a recording was successfully stored locally within the listener device. At a later time, the client requested the audio data from the listener device through the assigned message. The listener successfully confirmed the receipt of this message and began the data transfer over the CEC network to the client as seen in Figure \ref{filetransfer}. The client successfully stored and deserialized this data into a valid file format. This further opens the possibility to a listener which could await keywords such as ``password'' passively or use voice-to-text technology to transfer days of conversations to an adversary.

\textbf{\attackthree} This attack was specified in order to demonstrate the concepts of Threat 3 (Facilitation of Attacks) and Threat 4 (Information Theft). In this local attack, the attacker uses \name to facilitate WPA/WPA2 handshake capture and prevent detection by a security system in place. In traditional handshake theft attacks, an attacker has to wait for a handshake to occur, this can take an indefinite amount of time as the WPA handshake is only transferred in specific cases \cite{wpasurvey}. If there is a time constraint, the attacker must attempt forced de-authentication \cite{wpacrack}. This raises the issue that forced de-authentication may be detected through a network scanner such as Wireshark or through more complex IDS \cite{deauthenticationdetection}. In this attack, we facilitate such a threat through the removal of time constraints.

\textit{Step 1 - Initial Configuration:} The attacker must be especially careful about the listener placement. The listener must be able to reach wireless network connections and must also come equipped with a wireless adapter capable of ``monitor mode'' for packet capture.  

\textit{Step 2 - Client Trigger:} The client triggers the listener's service wireless attack module. This activates the wireless adapter in monitor mode with airmon-ng, then begins the capture with airodump-ng using the wlan1 interface and BSSID ``7C:8B:CA:49:45:D2'' in the listener device. Airodump-ng process is opened in separate terminal using Python's \texttt{os} import command. This places the listener in a passive state which awaits handshakes to naturally occur without forced de-authentication. The attacker is not needed for the duration of this capture. At a later time, an authorized user connects and the handshake is captured passively.

\textit{Step 3 - Handshake Retrieval:} The attacker reconnects with the client and requests the handshake from listener. The listener first cleans the capture \texttt{.cap} file using \texttt{wpaclean}. This greatly reduces the file size and the transfer begins. The attacker can finally receive the cleaned capture through the \CEC File I/O module.

\begin{figure}[t]
\footnotesize
\centering{\includegraphics[width=0.45 \textwidth]{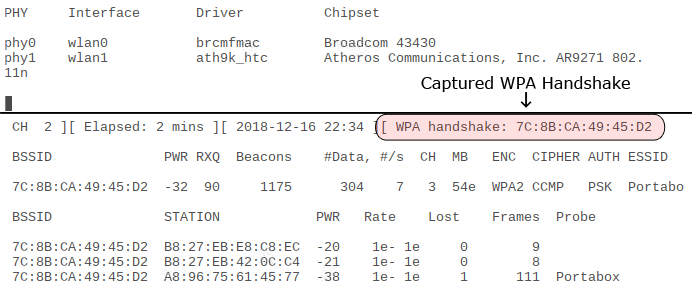}}
    \caption{Attack 3--Running handshake capture with Aircrack-ng.}
\label{handshake}
\vspace{-0.17in}
\end{figure} 

\textit{Evaluation:} Local \CEC client triggers for the activation of this attack proved entirely successful. Activation of the wireless module, Airodump-ng, and cleanup functions succeeded as seen in Figure \ref{handshake}. With the capture size reduced, the handshake was transferred to the local client successfully. This process would allow the attacker to retrieve the handshake at a later date and use more computing resources to attempt to crack the handshake and gain unauthorized access to the network. This would then allow the attacker to enable remote functionality to their own listener.

\textbf{\attackfour} This attack was developed to demonstrate Threat 5 (Denial of Service) through arbitrary sniffing and control of a device. In this attack, the attacker uses functionality from the Python-based listener service to target a specific device in the HDMI distribution. She also takes advantage of the nature of \CEC to sniff and detect when a device has been turned on. This attack can be divided into three main steps.

\textit{Step 1 - Activation:}  The listener awaits attack activation. It awaits commands either from the local client (through a walk) or from a remote client to activate the targeted attack. Once a command is received, the listener activates the attack. 

\textit{Step 2 - Sniffing:} The listener is set within an HDMI distribution and monitors \CEC packets flowing through the distribution. We particularly listen to the data commands ``84:00:00:00'', ``87:1f:00:08'' and ``80:00:00:30:00'' from any incoming source. These values, usually signify a device broadcasting to HDMI distribution devices that its power state has changed and has been turned on. More specifically, 84 reports physical address, 87 reports vendor id, and 80 reports a routing change. In this particular attack, the attacker targets a \CEC enabled display, the Sharp television.

\textit{Step 3 - DoS attack:} Once the attack is active the listener awaits commands associated with power state change within the HDMI distribution. Once the power state change is detected it sends the \CEC shutoff command ``20:36'' to the display (ID: 0) in the distribution. This automatically powers off the display as soon as it is powered on.

\begin{figure}[t]
\centering{\includegraphics[width=0.4 \textwidth]{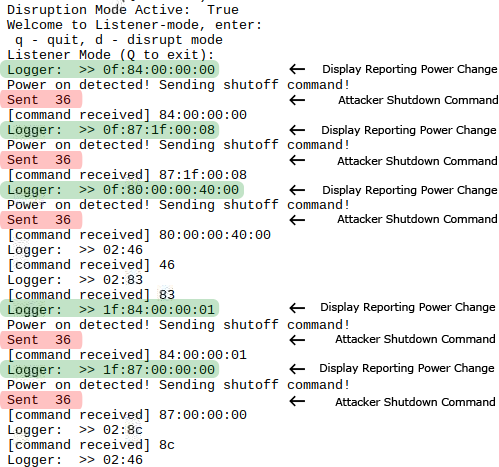}}
    \caption{Attack 4--TV Power state change and execution of targeted attack.}
\label{disrupt}
\vspace{-0.2in}
\end{figure}

\textit{Evaluation:} The listener began in an inactive state as expected with passive listening of the \CEC commands. Powering the display did not cause any changes in this inactive state. The listener successfully received the activation command over remote and local clients, activating the attack mode. With this mode active, the display was manually powered on. The module in our service successfully identified the power state change in the display and provided the shutoff command as seen in Figure \ref{disrupt}. The display received the shutoff commands and immediately powered off as expected. No matter which method of powering on, the attack could not be avoided, successfully executing the DoS attack. We additionally had another notable finding while performing this attack. That is, during DoS, the user was prevented from disabling CEC control within the system. Additionally, this attack may prove difficult to detect as it may be mistaken for a malfunctioning display.

\textbf{\attackfive} We developed this attack to demonstrate Threat 5 (Denial of Service) through broadcast functionality. This attack abuses the broadcast function in \CEC to cause a DoS condition in any display within a given HDMI distribution. This attack specifically targets displays by producing standard \CEC commands for source and input control. We divide this attack into three steps.

\textit{Step 1 - Insertion of Attacker Listener:} The listener device is placed in any location of the HDMI distribution. The device then awaits instructions from a client service to begin the attack. In the case of an available wireless connection, the listener's Remote Access Module becomes active.

\textit{Step 2 - Activation Phase:} The listener activates in two different methods: (1) the listener receives a direct command from a client service to begin the attack. (2) the listener receives through a remote client with the \texttt{DOS1} command.

\textit{Step 3 - DoS attack phase:} After activation conditions are reached, the listener device begins broadcast of various \textit{display input change} commands. These are standard \CEC commands accepted by enabled televisions to adjust the active source on the display device. The \CEC distribution is flooded with a broadcast loop: power on (``20:04''), input 1 (``82:10:00''), input 2 (``82:20:00''), input 3 (``82:30:00''), and input 4 (``82:40:00''). This renders the displays unusable by the user, effectively creating a DoS attack.

\begin{figure}[t]
\centering{\includegraphics[width=0.45 \textwidth]{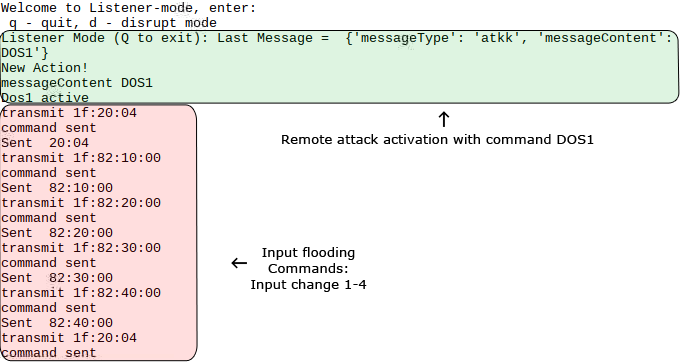}}
\vspace{-0.1in}
    \caption{Attack 5--Input-change induced DoS attack. Executed by remote attacker with command \texttt{DOS1}.}
\label{dos}
\vspace{-0.1in}
\end{figure}

\textit{Evaluation:} In this attack, the listener began in an idle state as intended in the distribution. The listener successfully received the activation command over remote and local clients. Then, it initiated the DoS broadcast loop over the entire distribution as depicted in Figure \ref{dos}. The attack first powered on the display if it was powered off. The loop then began rapid input change over all inputs on the display. The display began to flash rendering it unusable. We noticed faster switching between inputs than if compared with manual input change. Another effect of this condition is that it made it impossible for the user to alter any settings in the display to disable external control after activation. 

\noindent \textit{Summary and findings:}
During testing of \name attacks, we identified a vendor-specific vulnerability, and are currently coordinating to report this finding to the product's respective manufacturer. \name can identify specific device information to develop further attacks. We have proven arbitrary control over HDMI devices which could be used to an attacker's advantage. Also, we enabled control of the TV volume and Amplifier volume with devices in our testbed. This control is completely feasible in an HDMI distribution with the concepts of \name. We find these attacks critical as they occur over a medium without any form of security mechanisms or existing techniques for mitigation. Via Attack 4, we found that the input change control could become a viable form of a visual attack. With these functions, display input changes could be used to trigger seizures (e.g., television epilepsy) with the rapid flickering of a display switching between inputs \cite{epilepsy}. We also consider volume control to an Amplifier device. A remote attacker with the control of a distribution can easily adjust the volume of devices with \CEC commands. Extended playback at high volumes is known to damage sound equipment \cite{Speakerdamage}. An implementation of Attack 1 would first allow an attacker to infer room occupancy via power state. Combining this with Attack 4, the attacker could peak the volume output in a room when nobody is present and cause gradual damage to the sound system, which cause a notable financial cost to the user. Combination of \name and targeted device attacks such as Attack 4 could also allow a malicious person to assume control of menu functions in specific HDMI devices. This would allow the attacker to change menu settings, make purchases, or update firmware through device-tailored command sequences. With attackers in constant search for new vectors of attack, disruption, data leakage, behavioral leakage, and any type of information leakage could present catastrophic outcomes to an organization. A conference room while in confidential use can be a target to eavesdropping and handshake theft, giving attackers a chance to acquire passwords, access codes, and confidential information. In normal usage, inferring devices and disrupting functionality is possible and may present a threat which many users have not considered or anticipated.

\vspace{-0.1in}
\section{Discussion: Defense Mechanisms}

Although it is not within the scope and the aimed contributions of this work, we briefly discuss possible defense mechanisms for \name attacks.

\noindent\textit{Challenges of Defense Mechanisms:} One of the largest challenges with securing CEC revolves around purpose, implementation, and proliferation. CEC, being an established means of communication for A/V devices, has been widely deployed in its current form in billions of devices by different vendors. An endeavor to update the protocol with security mechanisms, or present a new version of the protocol has to consider discontinued devices, vendor-specific implementations, and backwards compatibility or face loss of functionality. Existing devices may not have the hardware, or connectivity capabilities to upgrade. All of these factors may make traditional security mechanisms too expensive, or impractical to deploy~\cite{magazine}. 

\noindent\textit{Removal of CEC:} An assumption of our attacks is that an incoming visitor has access to all \CEC features. This can be disabled through the complete removal of \CEC Pin 13 in the connection \cite{HDMIPinout}. CEC-less adapters/cables that partially or completely prevent the \CEC propagation within distribution may accomplish this goal.

\noindent\textit{Disable CEC via Setup:} Specific devices do not use CEC functionality due to external control via IP, Serial, IR, etc. It is a good practice to remove \CEC functionality in devices which do not require it. Many devices come with \CEC control enabled by default. Disabling CEC control may help mitigate DoS and any attack dependent on device control. However, it will not help against topology inference from devices which still provide their information after \CEC control is disabled. We found that some devices will still report address, and \CEC details on request even if \CEC control functionality is disabled. 

\noindent\textit{Awareness:} Knowing how devices operate, how they propagate, and strategic placement of CEC-less adapters/cabling may limit unauthorized access to \CEC and prevent CEC-based attacks. Awareness serves as a middle ground between fully removing \CEC and completely allowing communication. For example, a guest presenter should not have control to the distribution beyond displaying his/her laptop, thus there is no need for the presentation podium to allow \CEC communication.

\noindent\textit{HDMI-Enabled IDS:} Preventing an unauthorized device from joining an HDMI distribution could be the best step towards complete HDMI-Based attack mitigation. A promising solution in this can be to design an IDS for \CEC as a preventive measure for \CEC based attacks, moving from manual device scanning to more automated approaches. Machine learning-based approaches that consider the state of different features in hardware distribution have proven to be very effective in other research works~\cite{iotdots, six}. However, existing solutions make use of code instrumentation~\cite{saint} or behavioral analysis~\cite{icc} to capture real-time data from systems~\cite{saint}, which may be difficult to implement in close-source environments like vendor-specific CEC firmware. 
\vspace{-0.1in}

\section{Conclusions} \label{sec:conclusions}
Today there are close to 10 billion High Definition Multimedia Interface (HDMI) devices in the world and HDMI has become the de-facto standard for the distribution of A/V signals in smart homes, office spaces, sports events, etc. A component \luisnew{of this widely-deployed interface }is the CEC protocol, which is used for HDMI control interactions. With no currently known security solutions in place or security implementations in the CEC protocol design, \CEC opens a realm of possibilities to attackers. In this work, we introduced a novel attack surface for HDMI distribution networks called \name and its proof-of-concept attacks on deployed HDMI systems. To the best of our knowledge, this is the first work that solely investigated the security of \name based attacks in HDMI distribution networks. Using \name, we analyzed the \CEC propagation and implemented a series of local and remote \CEC based attacks as a proof-of-concept design. Specifically, we implemented malicious device analysis, eavesdropping, Denial of Service, targeted device attacks, and facilitation of existing attacks using CEC. \name presents a critical threat as these attacks cannot be detected or mitigated through traditional means of network protection. We further evaluated these novel attacks and highlighted their implications, including arbitrary control over CEC-enabled devices through a distribution. This work, aims to shed light on vulnerabilities in the largely deployed HDMI distribution. Consequences of these vulnerabilities can largely impact all users of HDMI and the security of any dependent system. Further, we discussed defense mechanisms to provide impactful and comprehensive security suggestions specific to the CEC protocol. As future work, we aim to develop a specific security mechanism for \name attacks, such as the design of an Intrusion Detection System.

\vspace{-0.05in}

\section*{Acknowledgments}

This work is partially supported by the US National Science Foundation (Awards: NSF-CAREER-CNS-1453647, NSF-1663051) and Florida Center for Cybersecurity's Capacity Building Program. The views are those of the authors only.

%
\bibliographystyle{ACM-Reference-Format}
\bibliography{references}


\begin{thebibliography}{26}


\ifx \showCODEN    \undefined \def \showCODEN     #1{\unskip}     \fi
\ifx \showDOI      \undefined \def \showDOI       #1{#1}\fi
\ifx \showISBNx    \undefined \def \showISBNx     #1{\unskip}     \fi
\ifx \showISBNxiii \undefined \def \showISBNxiii  #1{\unskip}     \fi
\ifx \showISSN     \undefined \def \showISSN      #1{\unskip}     \fi
\ifx \showLCCN     \undefined \def \showLCCN      #1{\unskip}     \fi
\ifx \shownote     \undefined \def \shownote      #1{#1}          \fi
\ifx \showarticletitle \undefined \def \showarticletitle #1{#1}   \fi
\ifx \showURL      \undefined \def \showURL       {\relax}        \fi
\providecommand\bibfield[2]{#2}
\providecommand\bibinfo[2]{#2}
\providecommand\natexlab[1]{#1}
\providecommand\showeprint[2][]{arXiv:#2}

\bibitem[\protect\citeauthoryear{{Aksu}, {Babun}, {Conti}, {Tolomei}, and
  {Uluagac}}{{Aksu} et~al\mbox{.}}{2018}]%
        {magazine}
\bibfield{author}{\bibinfo{person}{H. {Aksu}}, \bibinfo{person}{L. {Babun}},
  \bibinfo{person}{M. {Conti}}, \bibinfo{person}{G. {Tolomei}}, {and}
  \bibinfo{person}{A.~S. {Uluagac}}.} \bibinfo{year}{2018}\natexlab{}.
\newblock \showarticletitle{Advertising in the IoT Era: Vision and Challenges}.
\newblock \bibinfo{journal}{\emph{IEEE Communications Magazine}}
  \bibinfo{volume}{56}, \bibinfo{number}{11} (\bibinfo{date}{November}
  \bibinfo{year}{2018}), \bibinfo{pages}{138--144}.
\newblock
\showISSN{0163-6804}
\urldef\tempurl%
\url{https://doi.org/10.1109/MCOM.2017.1700871}
\showDOI{\tempurl}


\bibitem[\protect\citeauthoryear{{Babun}, {Aksu}, and {Uluagac}}{{Babun}
  et~al\mbox{.}}{2017}]%
        {icc}
\bibfield{author}{\bibinfo{person}{L. {Babun}}, \bibinfo{person}{H. {Aksu}},
  {and} \bibinfo{person}{A.~S. {Uluagac}}.} \bibinfo{year}{2017}\natexlab{}.
\newblock \showarticletitle{Identifying counterfeit smart grid devices: A
  lightweight system level framework}. In \bibinfo{booktitle}{\emph{2017 IEEE
  International Conference on Communications (ICC)}}. \bibinfo{pages}{1--6}.
\newblock
\urldef\tempurl%
\url{https://doi.org/10.1109/ICC.2017.7996877}
\showDOI{\tempurl}


\bibitem[\protect\citeauthoryear{Babun, Sikder, Acar, and Uluagac}{Babun
  et~al\mbox{.}}{2018}]%
        {iotdots}
\bibfield{author}{\bibinfo{person}{Leonardo Babun}, \bibinfo{person}{Amit~Kumar
  Sikder}, \bibinfo{person}{Abbas Acar}, {and} \bibinfo{person}{A.~Selcuk
  Uluagac}.} \bibinfo{year}{2018}\natexlab{}.
\newblock \bibinfo{title}{IoTDots: A Digital Forensics Framework for Smart
  Environments}.
\newblock
\newblock
\showeprint[arxiv]{cs.CR/1809.00745}
\urldef\tempurl%
\url{https://arxiv.org/pdf/1809.00745.pdf}
\showURL{%
\tempurl}


\bibitem[\protect\citeauthoryear{Baharudin, Ali, Darus, and Awang}{Baharudin
  et~al\mbox{.}}{2015}]%
        {deauthenticationdetection}
\bibfield{author}{\bibinfo{person}{N. Baharudin}, \bibinfo{person}{F.~H.~M.
  Ali}, \bibinfo{person}{M.~Y. Darus}, {and} \bibinfo{person}{N. Awang}.}
  \bibinfo{year}{2015}\natexlab{}.
\newblock \showarticletitle{Wireless Intruder Detection System (WIDS) in
  Detecting De-Authentication and Disassociation Attacks in IEEE 802.11}. In
  \bibinfo{booktitle}{\emph{2015 5th International Conference on IT Convergence
  and Security (ICITCS)}}. \bibinfo{pages}{1--5}.
\newblock
\urldef\tempurl%
\url{https://doi.org/10.1109/ICITCS.2015.7293037}
\showDOI{\tempurl}


\bibitem[\protect\citeauthoryear{Celik, Babun, Sikder, Aksu, Tan, McDaniel, and
  Uluagac}{Celik et~al\mbox{.}}{2018}]%
        {saint}
\bibfield{author}{\bibinfo{person}{Z.~Berkay Celik}, \bibinfo{person}{Leonardo
  Babun}, \bibinfo{person}{Amit~Kumar Sikder}, \bibinfo{person}{Hidayet Aksu},
  \bibinfo{person}{Gang Tan}, \bibinfo{person}{Patrick McDaniel}, {and}
  \bibinfo{person}{A.~Selcuk Uluagac}.} \bibinfo{year}{2018}\natexlab{}.
\newblock \showarticletitle{Sensitive Information Tracking in Commodity IoT}.
  In \bibinfo{booktitle}{\emph{27th {USENIX} Security Symposium ({USENIX}
  Security 18)}}. \bibinfo{publisher}{{USENIX} Association},
  \bibinfo{address}{Baltimore, MD}, \bibinfo{pages}{1687--1704}.
\newblock
\showISBNx{978-1-931971-46-1}
\urldef\tempurl%
\url{https://www.usenix.org/conference/usenixsecurity18/presentation/celik}
\showURL{%
\tempurl}


\bibitem[\protect\citeauthoryear{Davis}{Davis}{2013}]%
        {nccgroup}
\bibfield{author}{\bibinfo{person}{Andy Davis}.} \bibinfo{year}{Aug,
  2013}\natexlab{}.
\newblock \bibinfo{title}{{What the HEC? Security implications of HDMI Ethernet
  Channel and other related protocols}}.
\newblock
\newblock
\urldef\tempurl%
\url{https://www.nccgroup.trust/globalassets/our-research/uk/whitepapers/2013/44con\_hdmi\_ethernet\_channel\_andy\\\_davis
  \_ncc\_group\_wp.pdf}
\showURL{%
\tempurl}


\bibitem[\protect\citeauthoryear{Davis}{Davis}{2012}]%
        {davisblackhat}
\bibfield{author}{\bibinfo{person}{Andy Davis}.} \bibinfo{year}{Mar,
  2012}\natexlab{}.
\newblock \bibinfo{title}{{HDMI : Hacking Displays Made Interesting}}.
\newblock
\newblock
\urldef\tempurl%
\url{https://media.blackhat.com/bh-eu-12/Davis/bh-eu-12-Davis-HDMI-Slides.pdf}
\showURL{%
\tempurl}


\bibitem[\protect\citeauthoryear{Dorsey}{Dorsey}{2017}]%
        {wpacrack}
\bibfield{author}{\bibinfo{person}{Brannon Dorsey}.} \bibinfo{year}{Jul,
  2017}\natexlab{}.
\newblock \bibinfo{title}{{Crack WPA/WPA2 Wi-Fi Routers with Aircrack-ng and
  Hashcat}}.
\newblock
\newblock
\urldef\tempurl%
\url{https://medium.com/@brannondorsey/crack-wpa-wpa2-wi-fi-routers-with-aircrack-ng-and-hashcat-a5a5d3ffea46}
\showURL{%
\tempurl}


\bibitem[\protect\citeauthoryear{{Google}}{{Google}}{2018}]%
        {tradenames}
\bibfield{author}{\bibinfo{person}{{Google}}.} \bibinfo{year}{2018}\natexlab{}.
\newblock \bibinfo{title}{{What is CEC?}}
\newblock
\newblock
\urldef\tempurl%
\url{{https://support.google.com/chromecast/answer/7199917?hl=en}}
\showURL{%
\tempurl}


\bibitem[\protect\citeauthoryear{Google}{Google}{2019}]%
        {AndroidCEC}
\bibfield{author}{\bibinfo{person}{Google}.} \bibinfo{year}{May,
  2019}\natexlab{}.
\newblock \bibinfo{title}{HDMI-CEC Control Service}.
\newblock
\newblock
\urldef\tempurl%
\url{https://source.android.com/devices/tv/hdmi-cec}
\showURL{%
\tempurl}


\bibitem[\protect\citeauthoryear{{HDMI Licensing LLC}}{{HDMI Licensing
  LLC}}{2018}]%
        {HDMIPinout}
\bibfield{author}{\bibinfo{person}{{HDMI Licensing LLC}}.}
  \bibinfo{year}{2018}\natexlab{}.
\newblock \bibinfo{title}{{Inside an HDMI Cable}}.
\newblock
\newblock
\newblock
\shownote{{https://www.hdmi.org/installers/ insidehdmicable.aspx}.}


\bibitem[\protect\citeauthoryear{{HDMI Licensing LLC}}{{HDMI Licensing
  LLC}}{2009}]%
        {HDMISpec}
\bibfield{author}{\bibinfo{person}{{HDMI Licensing LLC}}.} \bibinfo{year}{Jun,
  2009}\natexlab{}.
\newblock \bibinfo{title}{{HDMI Specification guide V1.4}}.
\newblock , \bibinfo{numpages}{pp. 213}~pages.
\newblock


\bibitem[\protect\citeauthoryear{Heyne}{Heyne}{2013}]%
        {Speakerdamage}
\bibfield{author}{\bibinfo{person}{Cliff Heyne}.} \bibinfo{year}{Jan,
  2013}\natexlab{}.
\newblock \bibinfo{title}{{AV Tip: How to Avoid Blowing Out Your Speakers}}.
\newblock
\newblock
\urldef\tempurl%
\url{https://www.audioholics.com/home-theater-connection/avoid-blowing-speakers}
\showURL{%
\tempurl}


\bibitem[\protect\citeauthoryear{Holman}{Holman}{2005}]%
        {HDMI}
\bibfield{author}{\bibinfo{person}{Kasey Holman}.} \bibinfo{year}{Dec,
  2005}\natexlab{}.
\newblock \bibinfo{title}{{HDMI Licensing\, LLC Announces Availability of HDMI
  1.2a Specification}}.
\newblock
\newblock
\urldef\tempurl%
\url{{https://www.hdmi.org/press/pr/pr\_20051227.aspx}}
\showURL{%
\tempurl}


\bibitem[\protect\citeauthoryear{Joseph I.~Sirven}{Joseph I.~Sirven}{2013}]%
        {epilepsy}
\bibfield{author}{\bibinfo{person}{MD Joseph I.~Sirven}.} \bibinfo{year}{Nov,
  2013}\natexlab{}.
\newblock \bibinfo{title}{{Photosensitivity and Seizures}}.
\newblock
\newblock
\urldef\tempurl%
\url{https://www.epilepsy.com/learn/triggers-seizures/photosensitivity-and-seizures}
\showURL{%
\tempurl}


\bibitem[\protect\citeauthoryear{{Kwikway}}{{Kwikway}}{[n.d.]}]%
        {HDMIBUS}
\bibfield{author}{\bibinfo{person}{{Kwikway}}.}
  \bibinfo{year}{[n.d.]}\natexlab{}.
\newblock \bibinfo{title}{{The HDMI-CEC Bus}}.
\newblock
\newblock
\urldef\tempurl%
\url{{http://wiki.kwikwai.com/index.php?title=The\_HDMI-CEC\_bus}}
\showURL{%
\tempurl}


\bibitem[\protect\citeauthoryear{Lashkari, Danesh, and Samadi}{Lashkari
  et~al\mbox{.}}{2009}]%
        {wpasurvey}
\bibfield{author}{\bibinfo{person}{Arash~Habibi Lashkari}, \bibinfo{person}{Mir
  Mohammad~Seyed Danesh}, {and} \bibinfo{person}{B. Samadi}.}
  \bibinfo{year}{2009}\natexlab{}.
\newblock \showarticletitle{A survey on wireless security protocols (WEP, WPA
  and WPA2/802.11i)}. In \bibinfo{booktitle}{\emph{2009 2nd IEEE International
  Conference on Computer Science and Information Technology}}.
  \bibinfo{pages}{48--52}.
\newblock
\urldef\tempurl%
\url{https://doi.org/10.1109/ICCSIT.2009.5234856}
\showDOI{\tempurl}


\bibitem[\protect\citeauthoryear{Niemietz, Somorovsky, Mainka, and
  Schwenk}{Niemietz et~al\mbox{.}}{2015}]%
        {NiemSmartTV}
\bibfield{author}{\bibinfo{person}{M. Niemietz}, \bibinfo{person}{J.
  Somorovsky}, \bibinfo{person}{C. Mainka}, {and} \bibinfo{person}{J.
  Schwenk}.} \bibinfo{year}{2015}\natexlab{}.
\newblock \showarticletitle{Not so Smart: On Smart TV Apps}. In
  \bibinfo{booktitle}{\emph{2015 International Workshop on Secure Internet of
  Things (SIoT)}}. \bibinfo{pages}{72--81}.
\newblock
\urldef\tempurl%
\url{https://doi.org/10.1109/SIOT.2015.13}
\showDOI{\tempurl}


\bibitem[\protect\citeauthoryear{Oren and Keromytis}{Oren and
  Keromytis}{2014}]%
        {OrenKeromytis}
\bibfield{author}{\bibinfo{person}{Yossef Oren} {and}
  \bibinfo{person}{Angelos~D. Keromytis}.} \bibinfo{year}{2014}\natexlab{}.
\newblock \showarticletitle{From the Aether to the
  Ethernet{\textemdash}Attacking the Internet using Broadcast Digital
  Television}. In \bibinfo{booktitle}{\emph{23rd {USENIX} Security Symposium
  ({USENIX} Security 14)}}. \bibinfo{publisher}{{USENIX} Association},
  \bibinfo{address}{San Diego, CA}, \bibinfo{pages}{353--368}.
\newblock
\showISBNx{978-1-931971-15-7}
\urldef\tempurl%
\url{https://www.usenix.org/conference/usenixsecurity14/technical-sessions/presentation/oren}
\showURL{%
\tempurl}


\bibitem[\protect\citeauthoryear{{Pulse-Eight}}{{Pulse-Eight}}{2018}]%
        {libcec}
\bibfield{author}{\bibinfo{person}{{Pulse-Eight}}.}
  \bibinfo{year}{2018}\natexlab{}.
\newblock \bibinfo{title}{{USB-CEC Adapter communication Library}}.
\newblock
\newblock
\urldef\tempurl%
\url{https://github.com/Pulse-Eight/libcec/}
\showURL{%
\tempurl}


\bibitem[\protect\citeauthoryear{Rondon}{Rondon}{[n.d.]}]%
        {myapi}
\bibfield{author}{\bibinfo{person}{L~Puche Rondon}.}
  \bibinfo{year}{[n.d.]}\natexlab{}.
\newblock \bibinfo{title}{"Remote HDMI-Walk Client"}.
\newblock
\newblock
\newblock
\shownote{http://phdapis-env.hzqhv6ugji.us-east-1.elasticbeanstalk.com/.}


\bibitem[\protect\citeauthoryear{Sikder, Aksu, and Uluagac}{Sikder
  et~al\mbox{.}}{2017}]%
        {six}
\bibfield{author}{\bibinfo{person}{Amit~Kumar Sikder}, \bibinfo{person}{Hidayet
  Aksu}, {and} \bibinfo{person}{A.~Selcuk Uluagac}.}
  \bibinfo{year}{2017}\natexlab{}.
\newblock \showarticletitle{6thSense: A Context-aware Sensor-based Attack
  Detector for Smart Devices}. In \bibinfo{booktitle}{\emph{26th {USENIX}
  Security Symposium ({USENIX} Security 17)}}. \bibinfo{publisher}{{USENIX}
  Association}, \bibinfo{address}{Vancouver, BC}, \bibinfo{pages}{397--414}.
\newblock
\showISBNx{978-1-931971-40-9}
\urldef\tempurl%
\url{https://www.usenix.org/conference/usenixsecurity17/technical-sessions/presentation/sikder}
\showURL{%
\tempurl}


\bibitem[\protect\citeauthoryear{Smith}{Smith}{2015}]%
        {smithcec}
\bibfield{author}{\bibinfo{person}{Joshua Smith}.} \bibinfo{year}{Nov,
  2015}\natexlab{}.
\newblock \bibinfo{title}{{High-Def Fuzzing : Exploring Vulnerabilities in
  HDMI-CEC}}.
\newblock
\newblock
\urldef\tempurl%
\url{https://media.defcon.org/}
\showURL{%
\tempurl}


\bibitem[\protect\citeauthoryear{Tsutsui}{Tsutsui}{2008}]%
        {hdmistandard}
\bibfield{author}{\bibinfo{person}{Akihiro Tsutsui}.}
  \bibinfo{year}{2008}\natexlab{}.
\newblock \showarticletitle{Latest trends in home networking technologies}.
\newblock \bibinfo{journal}{\emph{IEICE transactions on communications}}
  \bibinfo{volume}{91}, \bibinfo{number}{8} (\bibinfo{year}{2008}),
  \bibinfo{pages}{2470--2476}.
\newblock


\bibitem[\protect\citeauthoryear{Wright}{Wright}{2018}]%
        {HDMIBil2}
\bibfield{author}{\bibinfo{person}{Doug Wright}.} \bibinfo{year}{Jan,
  2018}\natexlab{}.
\newblock \bibinfo{title}{{Shipments of Products with HDMI Interface Nears 900
  Million Devices in 2017; Total Installed Base Approaches Seven Billion}}.
\newblock
\newblock


\bibitem[\protect\citeauthoryear{Zhang, Demetriou, Mi, Diao, Yuan, Zong, Qian,
  Wang, Chen, Tian, Gunter, Zhang, Tague, and Lin}{Zhang et~al\mbox{.}}{2017}]%
        {ZhangIoT}
\bibfield{author}{\bibinfo{person}{Nan Zhang}, \bibinfo{person}{Soteris
  Demetriou}, \bibinfo{person}{Xianghang Mi}, \bibinfo{person}{Wenrui Diao},
  \bibinfo{person}{Kan Yuan}, \bibinfo{person}{Peiyuan Zong},
  \bibinfo{person}{Feng Qian}, \bibinfo{person}{XiaoFeng Wang},
  \bibinfo{person}{Kai Chen}, \bibinfo{person}{Yuan Tian},
  \bibinfo{person}{Carl~A. Gunter}, \bibinfo{person}{Kehuan Zhang},
  \bibinfo{person}{Patrick Tague}, {and} \bibinfo{person}{Yue{-}Hsun Lin}.}
  \bibinfo{year}{2017}\natexlab{}.
\newblock \showarticletitle{Understanding IoT Security Through the Data Crystal
  Ball: Where We Are Now and Where We Are Going to Be}.
\newblock \bibinfo{journal}{\emph{CoRR}}  \bibinfo{volume}{abs/1703.09809}
  (\bibinfo{year}{2017}).
\newblock
\showeprint[arxiv]{1703.09809}
\urldef\tempurl%
\url{http://arxiv.org/abs/1703.09809}
\showURL{%
\tempurl}


\end{thebibliography}

\end{document}